\documentclass[9pt,twocolumn,twoside]{osajnl}

\journal{ol} 

\setboolean{shortarticle}{false} 

\ifthenelse{\boolean{shortarticle}}{\colorlet{color2}{color2b}}{\colorlet{color2}{color2}} 

\title{Tailorable Dispersion in a Four-Wave Mixing Laser}

\author[1]{Demetrious T. Kutzke}
\author[1]{Owen Wolfe}
\author[2]{Simon M. Rochester}
\author[2,3,4]{Dmitry Budker}
\author[1]{Irina Novikova}
\author[1,*]{Eugeniy E. Mikhailov}

\affil[1]{Department of Physics, College of William $\&$ Mary, Williamsburg, Virginia 23187, USA}
\affil[2]{Rochester Scientific, LLC, El Cerrito, CA, 94530, USA}
\affil[3]{Department of Physics, University of California, Berkeley, CA 94720-7300, USA}
\affil[4]{Helmholtz Institute Mainz, Johannes Gutenberg University, 55099 Mainz, Germany}

\affil[*]{Corresponding author: eemikh@wm.edu}

\dates{Compiled \today}

\ociscodes{(270.1670) Coherent optical effects; (270.4180) Multiphoton processes; (270.3430) Laser theory.}

\doi{\url{http://dx.doi.org/10.1364/ao.XX.XXXXXX}}

\begin{abstract}
We present experimental results demonstrating controllable dispersion in a ring laser
by monitoring the lasing-frequency response to cavity-length variations. Pumping on an
N-type level configuration in ${}^{87}$Rb, we tailor the intra-cavity dispersion slope
by varying experimental parameters such as pump-laser frequency, atomic density,
and pump power. As a result, we can tune the 
pulling factor (PF), i.e.\ the ratio of the laser frequency shift to the empty cavity 
frequency shift, of our laser by more than an order of magnitude.
\end{abstract}

\setboolean{displaycopyright}{true}

\begin{document}

\maketitle
\thispagestyle{fancy}

\ifthenelse{\boolean{shortarticle}}{\ifthenelse{\boolean{singlecolumn}}{\abscontentformatted}{\abscontent}}{}


The sensitivity of a cavity's resonant frequency to changes in its optical path length is at the heart of many interferometric optical
measurements~\cite{Siegman_book}. This effect is employed in lasers to tune their operating
frequency, and in precision path-length measurements to permit a distance
measurement resolution at the sub-atomic-length scale (as required for
the detection of gravitational waves~\cite{gwFirstDetection2016prl}).
State-of-the-art high-finesse cavities are capable of seeing a frequency shift due to one excited atom
in the cavity, providing the basis for cavity-quantum-electrodynamic
tests~\cite{kimble1998prl_single_atom_cavity_qed}.
The same effect, employed in optical gyroscopes, allows detection of
minuscule changes in rotation speed
~\cite{stedman97gyros, velikoseltsev2014QE_lage_gyros}.
 
All of the above is reflected in the standard textbook formula governing the
frequency shift $\Delta f_\text{empty}$ of an empty cavity with respect to a change in its
round-trip path length $p$: $\Delta f_\text{empty} = - f_0 \frac{\Delta p}{p}$, where $f_0$ is the resonant frequency for the path length $p$.
However, if there is a medium inside the cavity, then the
cavity response must account for the dispersion of the
medium~\cite{shahriar2007pra_fast_gyro}:
\begin{equation}
	\label{eq:disp_cavity_shift}
	\Delta f_\text{dispersive} 
	= \Delta f_\text{empty} \frac{1}{n_0+f_0 \frac{\partial n}{\partial f}}
	= \Delta f_\text{empty} \frac{1}{n_g},
\end{equation}
where $n_g$ is the group refractive index,
$n_0$ is the refractive index at the light frequency $f_0$, and
$\frac{\partial n}{\partial f}$ is the dispersion.

If the dispersion is
positive and sufficiently large, then $n_g$ will be $\ge 1$. This
suppresses the cavity sensitivity to length perturbations. 
The regime with $n_g \ge 1$ is called
the subluminal or the slow-light regime, since the group velocity $v_g = c/n_g < c$.
This regime is useful in producing
stable laser systems, where the laser frequency is immune to cavity-length fluctuations~\cite{sabooni2013prl_linenarrowing_slow_light}.

If the dispersion is
negative and sufficiently large, then $n_g$ will be $\le 1$.
This is known as
the superluminal or the fast-light regime, since the group velocity 
exceeds the speed of light in vacuum. 
In this regime the cavity response to the path change is enhanced. 
For the extreme case of $n_g = 0$, \eqref{eq:disp_cavity_shift}
seems to indicate that the resonance frequency has an infinite response to a cavity path-length change; however, an infinite number of round trips are required to
achieve the steady state~\cite{Smith2008PRA}.
Nevertheless, a quite significant enhancement
of the cavity response can
theoretically be achieved, as suggested for cases of
passive cavities filled 
with dispersive media~\cite{Myneni2012PRA,Smith2009PRA} and cavities with actively
controlled atomic dispersion~\cite{shahriar2005FiO,
shahriar2007pra_fast_gyro}. For the passive-cavity case, there has been experimental
demonstration of increasing cavity response by a factor of
5~\cite{Myneni2012PRA}.

So far there has been no direct demonstration of the {\em dispersion-modified laser-frequency}
dependence on the cavity path change.
The goal of this paper is to directly show 
the tailorable intra-cavity
dispersion of the lasing cavity via its modified response.
We demonstrate that the cavity
response can be changed by more than an order of magnitude by varying experimental parameters.


Maintaining the fast-light regime for a lasing field is challenging. Due
to the Kramers-Kronig relationship, an amplification line has positive
dispersion, which is associated with the slow-light regime. To circumvent this,
there must be a local absorption dip in the overall laser gain line.

\begin{figure}[h]
	\includegraphics[width=0.7\columnwidth]{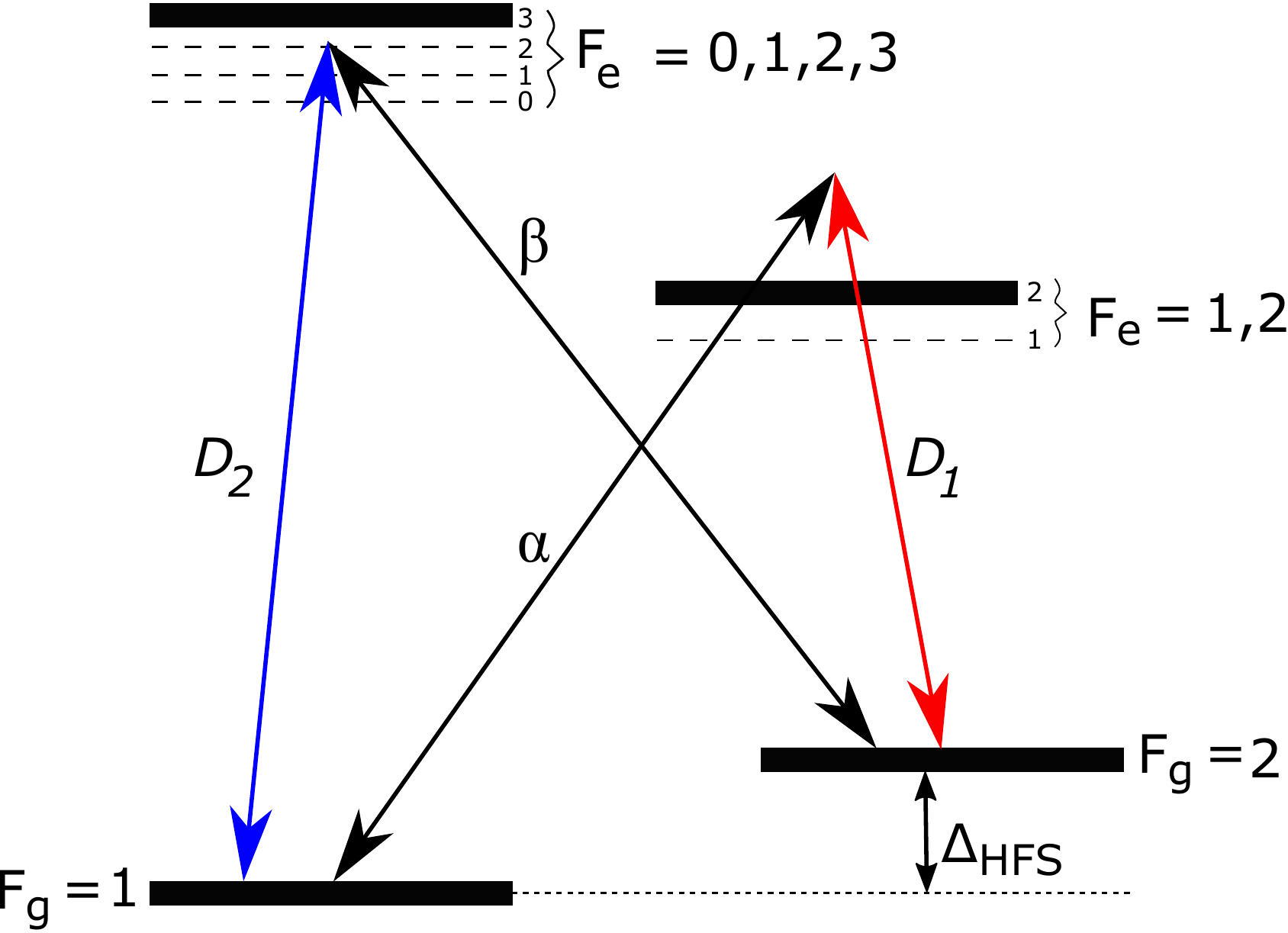}
	\caption{(Color online) The four-wave mixing  pumping N-scheme and
		relevant hyperfine levels of $^{87}$Rb.}
	\label{fig:n_scheme}
\end{figure}

We base our setup on an N-configuration pumping scheme outlined
in~\cite{mikhailov2013jmo_fast_N_scheme}. In this scheme, two 
strong pump lasers are tuned in the vicinity of the $D_1$ and the $D_2$ Rb lines as shown in
Fig.~\ref{fig:n_scheme}. This creates favorable conditions for four-wave
mixing of the pumps and the two additional new fields $\alpha$ and $\beta$
that result from the Raman-gain conditions.
The $D_1$ pump field by itself creates subluminal conditions for the $\alpha$ field. In the
presence of the strong $D_2$ pump, however, the gain line of the $\alpha$ 
field splits, which creates the dip necessary for the
fast-light regime. The $D_2$ pump also increases the gain for $\alpha$
via the four-wave mixing process.
During our preliminary studies on the N-scheme~\cite{mikhailovOE2014fwm_in_ring_cavity}, we
observed the fast-light regime without a cavity and also demonstrated
that lasing is possible in this scheme.

There are
several other proposals for achieving lasing in the fast-light regime. For example, a proposal from the Shahriar
group \cite{shahriar2007prl_wlc_demo} suggests that the presence of two Raman gain lines in the vicinity of each other
will provide an absorption-like feature between the gain lines while retaining the positive gain necessary for lasing. For another design from this group, in which
an absorptive dip is placed on top of a single Raman gain
line~\cite{shahriar2015OE_towards_superl_laser_with_dip}, indirect
measurements suggest an achievable cavity-response enhancement by a factor of 190.


\begin{figure}
	\includegraphics[width=1.0\columnwidth]{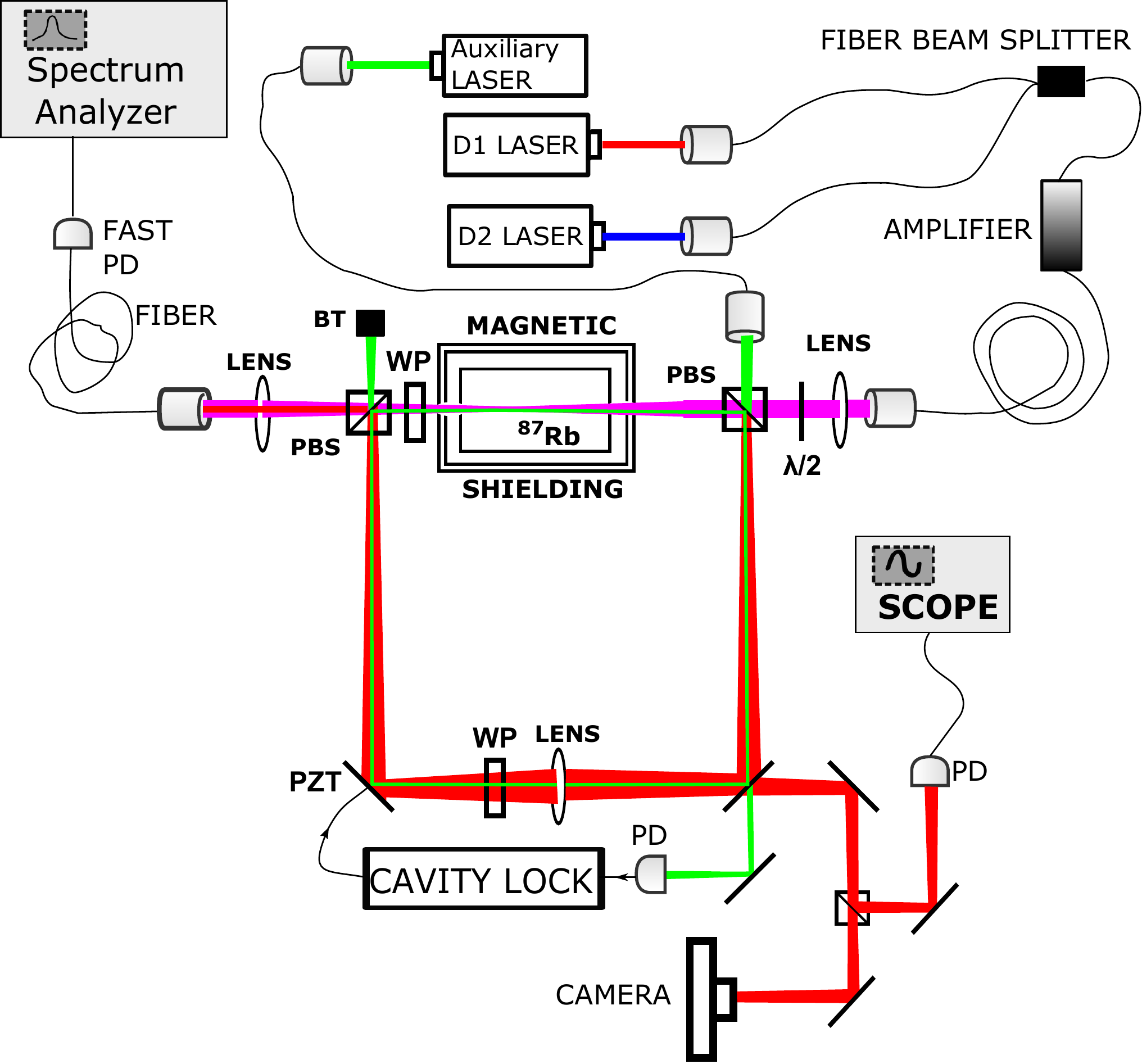}
	\caption{Schematic of the setup.}
	\label{fig:setup}
\end{figure}

A schematic representation of our experimental setup is shown in Fig.~\ref{fig:setup}.
The gyroscope is a square-shaped cavity with perimeter $p=80$~cm, into which 
we inject two pump-laser fields 
tuned to the ${}^{87}$Rb $D_{1}$ line ($795$~nm) and $D_{2}$ line (780 nm),
respectively. These pumps are first combined using a fiber-optic beam splitter and
injected into a solid-state amplifier to boost their total power level to
about 200~mW  before injecting them into the cavity.
The power ratio of the pumps was approximately 1 for all of the data presented here.
A cylindrical cell (length 22~mm, diameter 
25~mm) filled with isotopically enriched ${}^{87}$Rb vapor and 5~torr of Ne buffer gas is placed inside of a 3-layer magnetic shielding and mounted inside the cavity. 
Two highly reflective flat mirrors and two polarizing
beam splitter (PBS) cubes comprise the cavity, as depicted in Fig.~\ref{fig:setup}. To stabilize
the cavity, we use an intracavity lens with a focal length
of 30~cm. Such a configuration effectively directs the most
focused part of the cavity mode inside of the Rb cell. The overall finesse of the empty cavity was about 30--40, largely determined by absorption by the Rb-cell windows ($\approx$~0.9--0.95) and intentional polarization distortion by two waveplates (WP), as discussed below.

To control
the cavity length, we lock
it to an auxiliary laser that is far detuned from any atomic
transitions and thus ``sees'' the empty cavity, unaffected by the atomic dispersion. The light from this laser is counterpropagating to the pumps' direction to avoid
saturation of the output detectors. 
While most of the auxiliary field passes through a polarizing beam splitter (PBS) and hits the beam
trap (BT), a small polarization rotation, introduced by a
waveplate, transfers a fraction of the auxiliary field into a cavity mode.
When locked, the cavity length follows the frequency change of the auxiliary laser ($\Delta
f_\text{empty}$).

The pump light's polarization is set so that it can pass through the beam splitters, so the pump fields do not circulate in the cavity. However, the fields $\alpha$ and
$\beta$ are produced with polarization orthogonal to that of the pump, and thus are reflected by the PBSs and can lase when cavity resonant conditions are met. 
%
In our setup, the gain line is  narrower than the cavity free
spectral range, so we have to tune the cavity frequency to match the atomic
gain line. Once the cavity detuning is such that it is lasing, we can measure the
lasing frequency. This is done by observing the beat note between the $D_1$
pump and the lasing field on a fast photodiode. The photodiode mostly
sees the pump fields passing through the output-cavity PBS. Normally,
the lasing field would be fully reflected by this PBS back into the cavity.
To allow a small pick-off, we rotate the light polarization by a few degrees using a waveplate between the atomic cell and the
output PBS, 
so that a small
fraction of the lasing field escapes the cavity in the same polarization as
pumps. We can thus observe the beat note between the newly generated field and the corresponding pump occurring at a frequency near the $^{87}$Rb ground level splitting ($6.8$~GHz). The
linewidth of each of the beat notes is limited by the spectral-analyzer
resolution bandwidth which was typically set to 300~kHz. While this particular detection scheme does not allow us to discriminate between the
beat note of the lasing field $\alpha$ and $D_1$ field or the $\beta$ field and
$D_2$, the theoretical analysis strongly indicates that the $\alpha$ field
has stronger gain and thus is the one which lases. If we block either of the pumps, the
lasing ceases. 

\begin{figure}
\includegraphics[width=0.9\columnwidth]{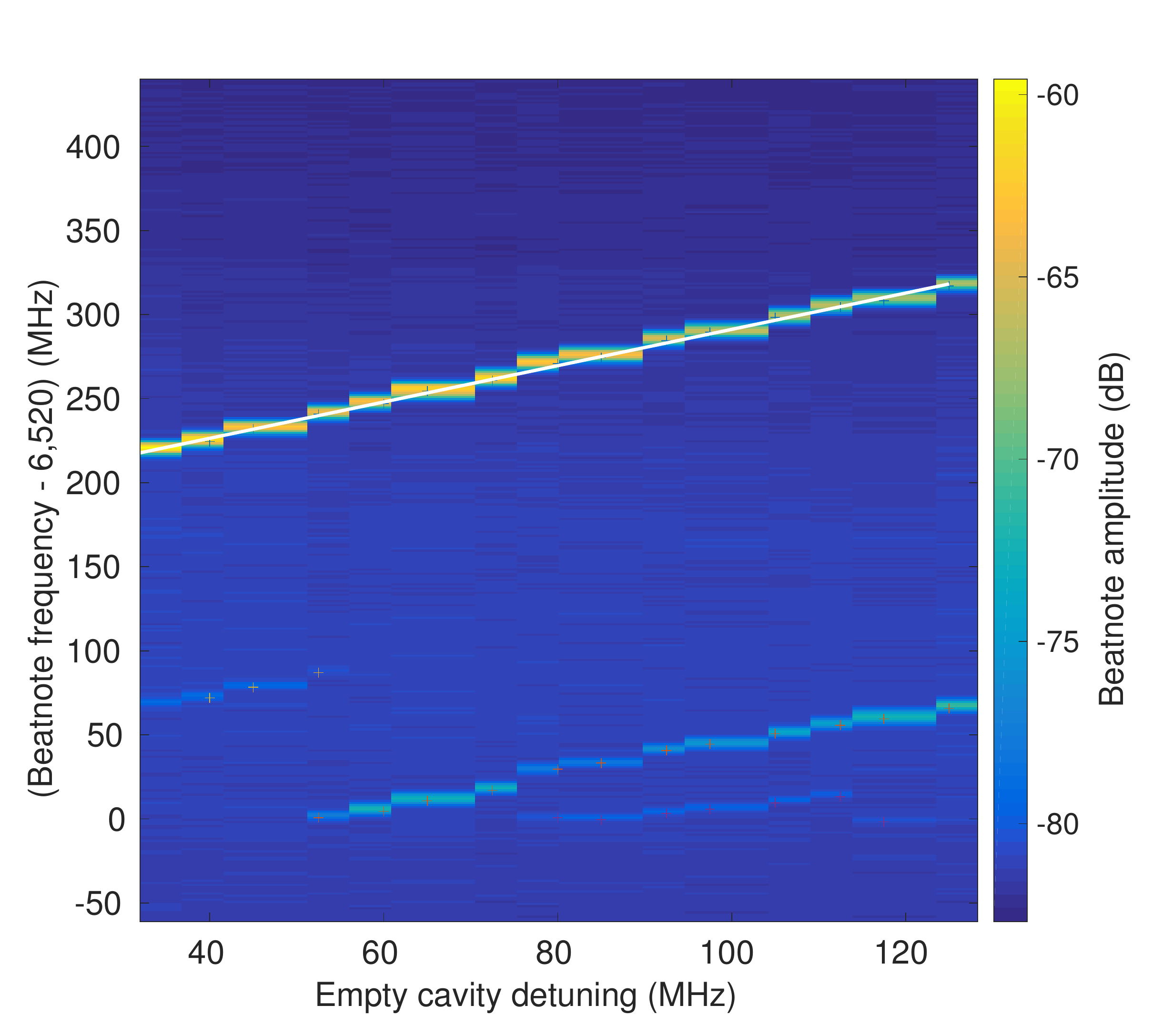}
\caption{
	(Color online)
	The beat-note-spectrum dependence on the empty-cavity detuning
	(cavity length change).
	The line demonstrates the linear fit used for the pulling factor
	extraction. Tracks of several competing spatial lasing modes are
	shown. Note that the tracks have different slopes (pulling factors).
	The $D_2$ pump was detuned 120 MHz toward lower frequencies from the 
	$F_g=2 \rightarrow F_{e} = 3$ transition. 
	The $D_1$ pump was detuned 1.5 GHz toward higher frequencies from the 
	$F_{g} = 1 \rightarrow F_{e} = 1$ transition.
	The combined $D_1$ and $D_2$ pump power was 230 mW.
}
\label{fig:beatnote_map_example}
\end{figure}

To measure the response of the generated-field frequency to cavity-length variations we continuously tune the frequency of the
auxiliary laser (to which the cavity is locked),  
thus changing the empty-cavity detuning ($\Delta f_\text{empty}$), and track the changes in the beat-note-frequency position. Since the pump frequency is held constant, the measured shift corresponds to the
lasing-frequency shift ($\Delta f_\text{dispersive}$). An example of this can be seen
in Fig.~\ref{fig:beatnote_map_example}. 
 Most of the time the dependence of the lasing detuning on the change in cavity length (and thus on $\Delta f_\text{empty}$) is linear.
We define the pulling factor (PF) as the slope of this linear dependence.
Combining this definition with \eqref{eq:disp_cavity_shift}, we obtain
\begin{equation}
	\label{eq:pulling_def}
	\text{PF} \equiv \frac{\Delta f_\text{dispersive}}{\Delta f_\text{empty}}=\frac{1}{n_{g}}.	
\end{equation}
If PF $>1$, the laser is in the fast-light regime with
enhanced response to changes in the cavity path length. Conversely,
for PF $<1$ the laser is less sensitive to fluctuations in cavity length.

Often we see more than
one beat note, corresponding to different spatial cavity-lasing modes as
observed by the camera. The different spatial lasing modes have different pulling factors (seen as separate lines with different slopes in 
Fig.~\ref{fig:beatnote_map_example}). For the results presented below we show only the mode that exhibited the greatest cavity response. Typically, this mode was the
TEM${}_{00}$ spatial lasing mode, which was usually 
the strongest.


\begin{figure}
	\includegraphics[width=0.8\columnwidth]{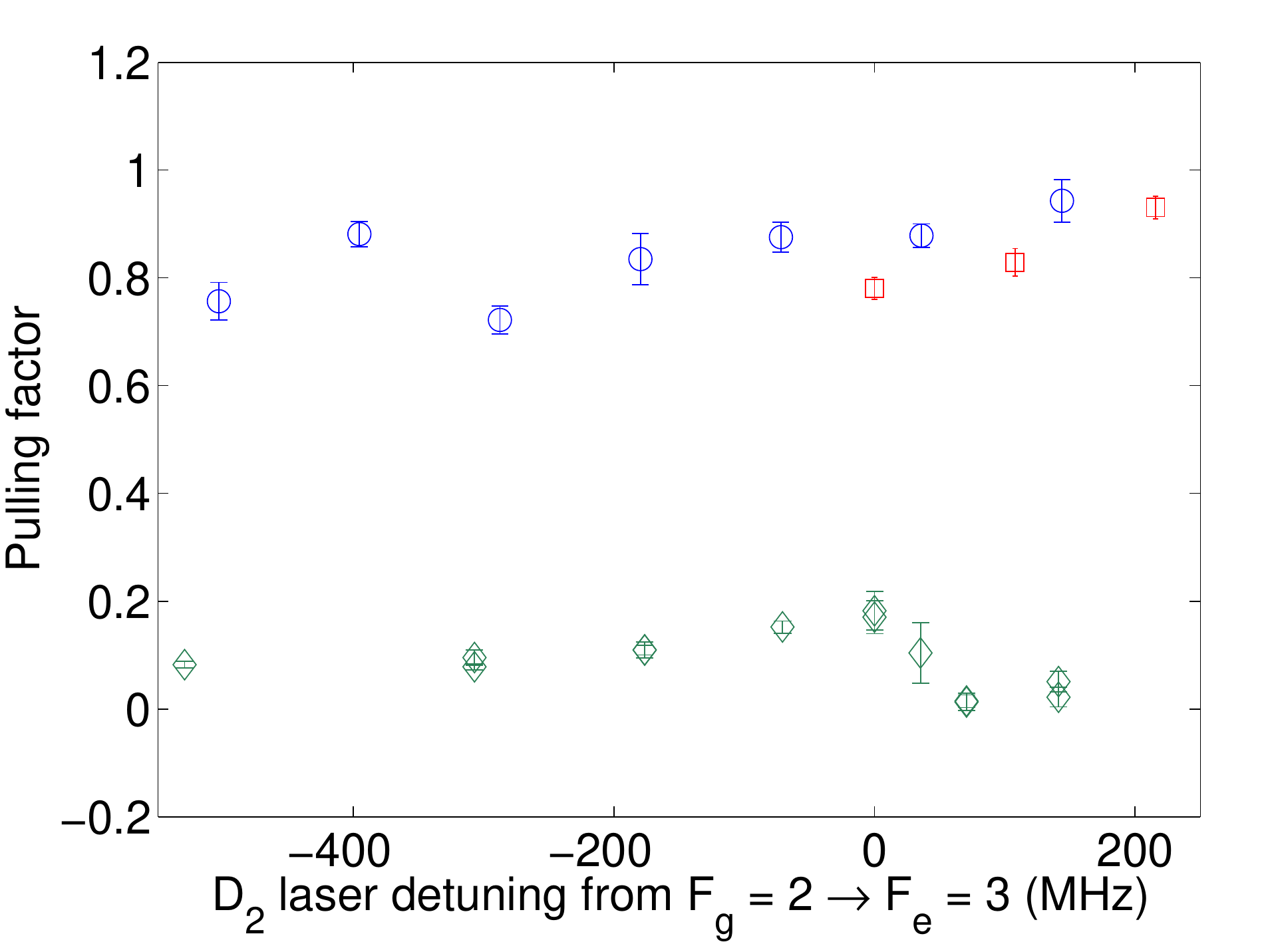}
	\caption{(Color online) PF vs. $D_2$ detuning data from the $F_{g} = 2 \rightarrow 
		F_{e} = 3$ transition. The blue circle markers indicate data
		collected at a total power of 210~mW,
		red squares at a total power of 178~mW,
		and green diamonds at a total power of 6~mW.
		Both the blue circle
		data and the red square data were collected at a fixed 
		Rb cell temperature of
		100~$^\circ$C. Likewise, 
		the green diamond markers indicate data collected at a  
		temperature of 80~$^\circ$C.
		All measurements were taken at a fixed $D_1$ detuning of 1.5 GHz towards
		higher frequencies
		from the $F_{g} = 1 \rightarrow F_{e} = 1$ transition.
		The error bars represent the uncertainty of the linear fit used
		for the PF extraction.
	}
	\label{fig:D2_detuning_PF_plot_1}
\end{figure}

Fig. \ref{fig:D2_detuning_PF_plot_1} presents measurements of the pulling factor as we varied the $D_2$
pump detuning from the Rb $F_{g} = 2 \rightarrow F_{e}=3$ transition, keeping the $D_1$ pump laser frequency
fixed at $+1.5$~GHz above
the $F_{g} = 1 \rightarrow F_{e} = 1$ transition.  We did
measurements in two different regimes. In the ``low-pump-power, low-atomic-density'' regime we set the total pump power at 6~mW, and maintained the Rb cell temperature at
80~$^\circ$C (corresponding to the atomic density $n=1.5 \times 10^{12}$~cm$^{-3}$). In the ``high-pump-power, high-atomic-density'' regime we increased the total pump power to either 208~mW or
178~mW, and were able then to operate at higher Rb cell temperature of 100~$^\circ$C
(atomic density $n=6.0 \times 10^{12}$~cm$^{-3}$). 

We see that in the high-pump-power regime,
there is a slight trend toward higher PF as we increase the $D_2$ pump frequency.
The PF reaches nearly unity as we approach approximately 200 MHz, at which point the lasing stops.
On the other hand, the low-pump-power regime exhibits a dispersion-like
dependence of the PF on the $D_2$ laser detuning. It reaches maximum when the $D_2$
detuning matches the $F_{g} = 2 \rightarrow F_{e} = 3$ transition and
drops to almost zero as we increase the detuning further. 
We note that the low-pump-power regime is associated with much
weaker lasing power, which is expected since the output power is related to
the pumping strength.

The drastic difference in PF for high and low power regimes
can be explained by the following. At low power the atomic gain line
is narrow since the power broadening of the gain line is small, thus the lasing frequency, which is dictated by the product of the cavity transmission and atomic
gain, is mainly governed by the atomic line position. 
On the other hand, in the high-power
regime, the power-broadened atomic gain line is wider than the cavity line, so
the lasing frequency mostly follows the cavity detuning.

\begin{figure}
	\includegraphics[width=0.8\columnwidth]{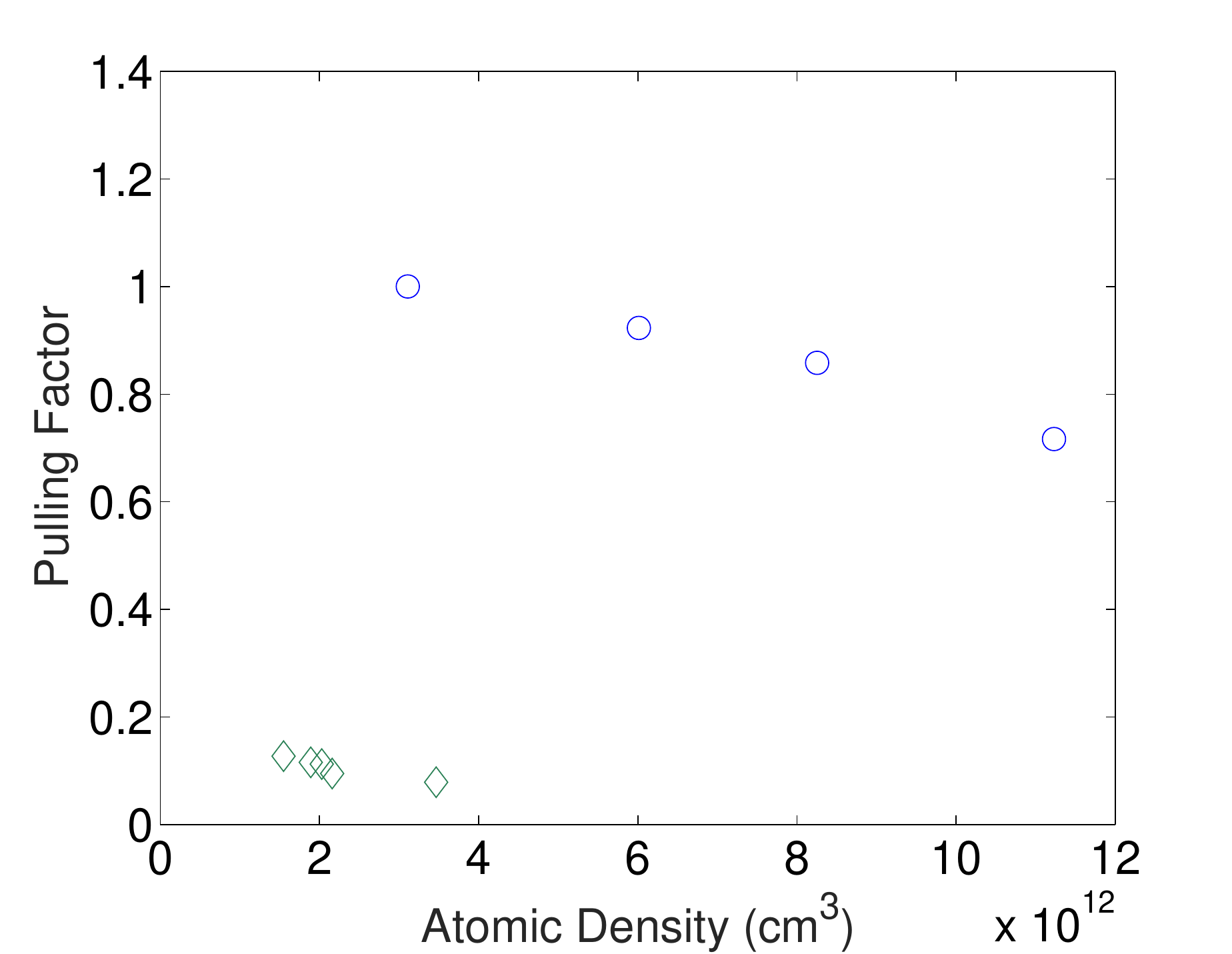}
	\caption{(color online) PF vs.\ atomic density. Both sets of measurements were taken 
	          at a fixed $D_1$ pump detuning of 1.5 GHz towards higher frequencies from the
		  $F_{g} = 1 \rightarrow F_{e} = 1$ transition. The blue circle data
		  were collected with a $D_2$ pump detuning of 110~MHz towards lower 
		  frequencies from the
                  $F_{g} = 2 \rightarrow F_{e} = 3$. The green diamond data were observed for a $D_2$
		  pump detuning of approximately 170~MHz towards lower frequencies from the 
		  $F_{g} = 2 \rightarrow F_{e} = 3$ transition. For the blue circle data the total pump
		  power was maintained at approximately 350~mW, while the green diamond data were
		  collected at approximately 7~mW.
	  }
	\label{fig:cell_dens_PF}
\end{figure}

Next, we studied the dependence of the PF on atomic density. The results
are shown in Fig.~\ref{fig:cell_dens_PF}. It is expected that with increased
atomic density, the dispersive-cavity properties will increasingly differ
from those of an empty cavity, and the PF will deviate from
unity. This behavior is very clear in
Fig.~\ref{fig:cell_dens_PF}: In the high-density and high-pump-power regime
the PF changes from 1 to approximately 0.7. In the low-power regime
the PF is small even in the low-density regime, but we also see a
decrease of the PF with the increase of atomic density.

\begin{figure}
	\includegraphics[width=1.0\columnwidth]{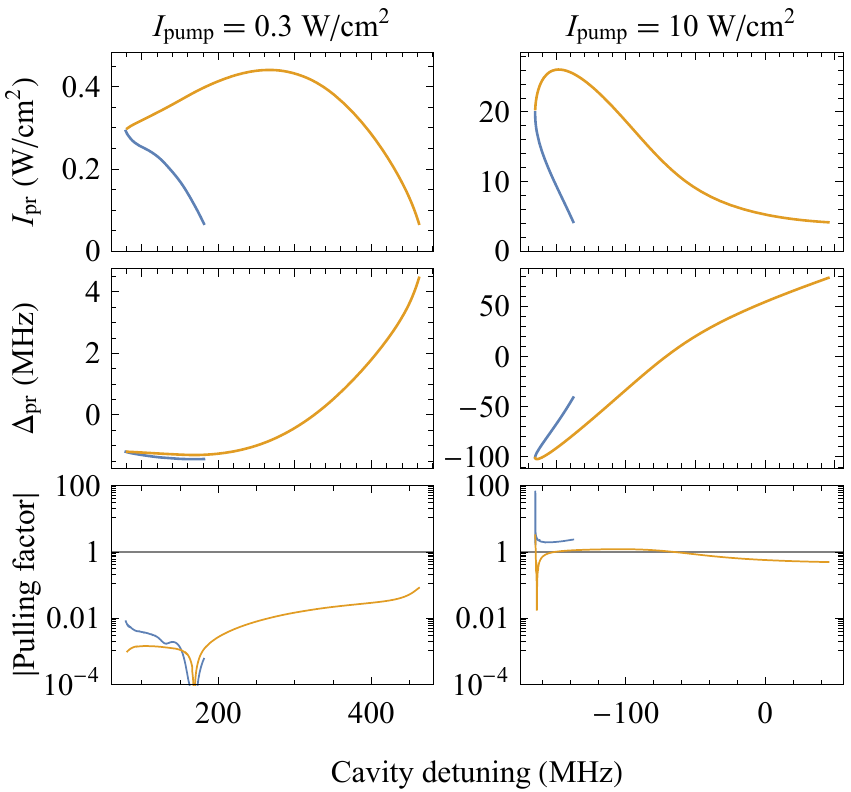}
	\caption{(color online) Numerical simulation results for intracavity probe power (top row), generated probe detuning (middle row) and absolute value of the pulling factor (bottom row) for the ``low'' ($300$~mW/cm$^2$) pump intensity  (left column) and for the ``high'' ($10^4$~mW/cm$^2$) pump intensity (right column).}
	\label{fig:num_calc}
\end{figure}

To better understand the prospects for extending the laser operation into the superluminal regime, we also conducted numerical simulations of probe-field generation for the four-level system shown in Fig.~\ref{fig:n_scheme}, including the effects of Doppler broadening. For a given set of pump parameters, the cavity round-trip complex amplitude gain for the probe light is calculated as a function of probe-field intensity and detuning, using the method described in Ref.~\cite{mikhailovOE2014fwm_in_ring_cavity}. For lasing to occur, we require unity round-trip gain; the probe intensities and detunings at which this happens are found as a function of cavity tuning. Fig.~\ref{fig:num_calc} shows results of the calculation with the $D_1$ pump tuned $1.5$~GHz above the optical resonance and the $D_2$ pump on resonance. The round-trip empty-cavity amplitude gain is set to 0.9 (corresponding to a finesse of 30) and the atomic density is $1.2\times 10^{12}~\mathrm{cm}^{-3}$ (lower than the experimental value to offset differences between the four-level system and the full hyperfine structure of Rb). Probe intensity $I_\mathrm{pr}$, detuning $\Delta_\mathrm{pr}$, and absolute value of the pulling factor (derivative of probe detuning with respect to cavity tuning) are shown for two pump intensities, roughly matching the experimental low- and high-pump-intensity regimes. Since there may be more than one combination of $I_\mathrm{pr}$ and $\Delta_\mathrm{pr}$ at which lasing can occur for each cavity tuning, different branches are represented by different-color lines.

One can see that the calculated behavior qualitatively matches the experimental observations. For all stable modes, the pulling factors are smaller than one for the most part, approaching unity at high pump intensity. However, even at lower pump intensity, there are certain places where the probe detuning vs.\ cavity tuning plot has very steep slope, resulting in a pulling factor much greater than one over a narrow cavity tuning range. Observing this regime experimentally will require more precise mode control than allowed by our current approach for measuring the pulling factor, and thus it is not surprising that we were not able to demonstrate this high pulling factor experimentally.   

In conclusion, we have shown that through a suitable choice of parameters one can change
the laser PF by more than an order of magnitude. 
While we were not able to demonstrate the fast-light regime in our lasing
system, we showed that we can bring the laser to the regime
where its sensitivity to cavity-length perturbations
is reduced to $0.014 \pm 0.012$ of its classical counterpart.
Such a reduced sensitivity regime is of interest for
the stable laser frequency systems used in metrology. 

We would like to thank Matt Simons and ShuangLi Du for help with initial setup
construction. We thank the Naval Air Warfare Center STTR program for supporting our research,
Contract No.\ N68335-13-C-0227.

\end{document}